\documentclass[twocolumn,showpacs,preprintnumbers,amsmath,amssymb]{revtex4}

\usepackage{graphicx} 
\usepackage{epsfig}
\usepackage{dcolumn}
\usepackage{bm}
\usepackage{wrapfig}

\begin{document}

\title{Robust control of individual nuclear spins in diamond}

\author{Benjamin Smeltzer}
\author{Jean McIntyre}
\author{Lilian Childress}
\affiliation{Department of Physics and Astronomy, Bates College, Lewiston ME}%

\date{\today}

\begin{abstract}
Isolated nuclear spins offer a promising building block for quantum information processing systems, but their weak interactions often impede preparation, control, and detection.  Hyperfine coupling to a proximal electronic spin can enhance each of these processes.  Using the electronic spin of the nitrogen-vacancy center as an intermediary, we demonstrate robust initialization, single-qubit manipulation, and direct optical readout of $^{13}$C, $^{15}$N, and $^{14}$N nuclear spins in diamond. These results pave the way for nitrogen nuclear spin based quantum information architectures in diamond.

\end{abstract}

\pacs{}
\maketitle

   The long coherence times of isolated nuclear spins have attracted considerable interest for applications in quantum information science \cite{Vandersypen04, Kane98, Baugh06, Witzel07,Morton08,Cappellaro09, Gorshkov09, Feng09}, offering the potential for devices exhibiting coherent evolution over timescales exceeding several seconds~\cite{Ladd05}. 
A significant challenge in such applications is development of techniques 
to prepare, manipulate, detect, and couple individual nuclear spins.  
Some of the most promising approaches use electronic spins as an intermediary to access single nuclear spins in solid state~\cite{Kane98, Baugh06, Witzel07, Morton08,Cappellaro09, Petta08, Bracker05, Verhulst05} or atomic~\cite{Feng09,Gorshkov09} systems.  While atomic systems enable optical preparation and near-perfect isolation of nuclear spins, nanoscale solid-state systems offer the possibility of manipulating and coupling spins on fast timescales.  Atom-like defects in solids~\cite{Wrachtrup06,Jelezko04b,Neumann08,Dutt07, Bala09, Mizuochi09} offer an intriguing interpolation between these approaches, allowing optical techniques in a setting amenable to fast control.   
 
We use the electronic spin of the negatively charged nitrogen-vacancy (NV), an atom-like defect, to access individual nuclear spins in the diamond lattice.  Even at room temperature, the NV spin is readily polarized and manipulated with optical and microwave fields~\cite{Manson06}, yet it can exhibit millisecond coherence times~\cite{Bala09, Mizuochi09}.  Moreover, its hyperfine interactions with 
nearby nuclear spins have allowed observation of single nuclear spin precession~\cite{Jelezko04b, Dutt07, Mizuochi09}, entanglement~\cite{Neumann08, Mizuochi09}, and two-bit quantum gates~\cite{Jelezko04b,Dutt07}.  As a result, the NV center has quickly become a strong candidate for spin-based quantum computation~\cite{Wrachtrup06} and communication~\cite{Childress06PRL} applications. 

Three nuclear spin species commonly interact with the NV center, offering different properties that lend themselves to different applications.  Experimental manipulation of single nuclear spins associated with the NV center has emphasized $I = \frac{1}{2}$ $^{13}$C~\cite{Jelezko04b,Childress06, Dutt07, Neumann08, Mizuochi09} and implanted $I = \frac{1}{2}$ $^{15}$N nuclei~\cite{Jacques09}, while magnetic resonance in $I = 1$ $^{14}$N  has been observed in ensemble experiments~\cite{He93b, Felton09}.   With the strongest hyperfine coupling, the nearest-neighbor $^{13}$C nuclear spins provide fast interactions but shorter coherence times~\cite{Dutt07, Mizuochi09}; like all $^{13}$C nuclei, they occur randomly as isotopic impurities in the lattice.  In contrast, the $^{14}$N and $^{15}$N species are weakly coupled, but they occur deterministically on the nitrogen of the NV center in natural ($^{14}$N) and isotope implanted ($^{15}$N) diamond~\cite{Fuchs08,Jacques09}. Scalable schemes -- based, for example, on arrays of dipole-coupled electron-nuclear spin pairs or optically connected quantum registers~\cite{Dutt07, Childress06PRL} -- are thus likely to favor the consistent properties of the nitrogen nuclear spins.  Moreover, the longest electron spin coherence times have been observed only in highly isotopically purified, unimplanted diamond~\cite{Bala09, Mizuochi09}, in which $^{13}$C and $^{15}$N spins are exceedingly rare.  
These considerations point towards the development of control techniques that will function for $^{15}$N and $^{14}$N as well as $^{13}$C nuclear spins.

In this paper, we demonstrate dynamic polarization, resonant manipulation, and direct measurement of single $^{13}$C, $^{14}$N, and $^{15}$N nuclear spins.  These techniques are robust in the sense that they do not rely on $I = \frac{1}{2}$ spin properties or weak (selective) electron spin resonance (ESR), and give rise to signals as strong as the ESR signal. Moreover, these techniques share a common origin in the nonsecular components of the hyperfine interaction between the NV electron spin and the nuclear spin.  The terms in the hyperfine interaction that give rise to electron-nuclear spin flip-flops allow robust polarization~\cite{Jacques09}, greatly enhance the Rabi frequency for nuclear magnetic resonance~\cite{Childress06}, and enable direct optical readout of the nuclear spin.  These mechanisms apply to all species of nuclear spin coupled to the NV center, and we show that it may be possible to extend them to work in arbitrary magnetic fields. 

In our experiments, NV centers in CVD-grown high-purity diamond (Sumitomo, natural isotopic abundance) are isolated using confocal microscopy, allowing selection of NV centers with different proximal nuclear spins. The centers are illuminated by 4mW of 532nm light for polarization and fluorescence detection, while Helmholtz coils and/or a permanent magnet are used to apply magnetic fields between 0 and 500G. Electron spin transitions are driven by resonant microwaves (MW) running through a 25 $\mu$m copper wire mounted on the sample; the same wire carries radio-frequency (RF) signals resonant with nuclear spin transitions.

The NV center in diamond has an electronic spin $S = 1$ in both its ground state (GS) and optically excited state (ES)~\cite{Manson06}; both orbital configurations exhibit a zero-field splitting ($\Delta = 2.87$ GHz (GS) or $D = 1.42$ GHz (ES)) between the $m_s = 0$ and $m_s = \pm 1$ spin projections along the NV axis ($\hat{z}$)~\cite{Fuchs08}.  The electronic structure of the NV center also includes at least one long-lived spin singlet state of intermediate energy that preferentially depopulates the ES $m_s = \pm 1$ states and decays into the GS $m_s = 0$ state.  The singlet state provides the mechanism for polarization and detection of the electronic spin state: under optical illumination, the $m_s = \pm 1$ states fluoresce more weakly than the $m_s = 0$ states before being polarized into the $m_s = 0$ state
~\cite{Manson06}.    

\begin{figure}[htbp] 
  \centering
  \includegraphics[width=3in,height=2in,keepaspectratio]{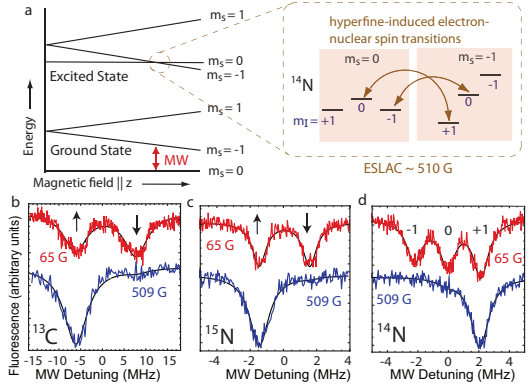}
  \label{fig:badfigure}
\caption{(a) Ground state (GS) and excited state (ES) spin sublevels of the NV center, showing the excited state level anti-crossing (ESLAC) near 510G. Inset shows hyperfine levels and transitions  for the $^{14}$N isotope.   
(b-d) ESR for an NV center coupled to a proximal nuclear spin in B=65 G(top) and 509 G (bottom). (b) a nearby $^{13}$C exhibits $P = 91(5)\%$ polarization; parentheses indicate the standard deviation statistical error in the final digit.  (c) $^{15}$N: P = 95(3)\%. (d)  $^{14}$N: P = 95(3)\%.}
\end{figure}

In certain magnetic fields, optical spin polarization can be extended to proximal nuclei~\cite{Fuchs08,Jacques09}.  Near 510G $|| \hat{z}$, the ES $m_s =-1$ and $m_s = 0$ states become degenerate (see Fig.~1a), allowing electron-nuclear spin flip-flops to occur near-resonantly; repeated cycling through the singlet state thus results in polarization into $m_s = 0$ with maximal $m_I$ (see Fig1b-d).  This mechanism was previously used to polarize $^{15}N$ and nearest-neighbor $^{13}$C nuclear spins~\cite{Jacques09} and also polarizes the $I = 1$ species $^{14}$N via the two-step process illustrated in Fig 1a.  Non-nearest-neighbor $^{13}$C nuclei may also be polarized in this manner (see Fig. 1b); however, not all $^{13}$C locations exhibit such polarization (data not shown) most likely because their hyperfine axes are not aligned in the ground and excited states~\cite{Felton09, Gali09}. 


\begin{figure}[htbp] 
  \centering
  \includegraphics[width=3.5in,height=3.5in,keepaspectratio]{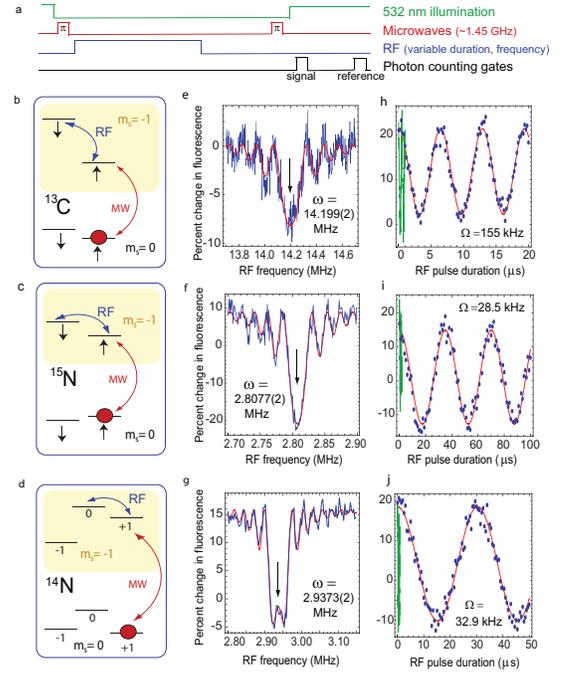}
  \label{fig:badfigure}
\caption{(a) Pulse sequence.   Following polarization, a MW $\pi$ pulse drives the system into the GS $m_s = -1$ state, where RF excitation excites nuclear spin transitions.  To measure the resulting population in the nuclear spin levels, a second MW $\pi$ pulse flips the electron spin back into $m_s = 0$, and subsequent fluorescence detection reveals the average nuclear spin projection. All data is taken in a magnetic field of 509G $||\hat{z}$. Data for $^{13}$C, $^{15}$N, and $^{14}$N is shown in (b,e,h), (c,f,i) and (d,g,j) respectively.  (b-d) Energy level diagrams for  hyperfine levels; arrows indicate transitions driven by RF and MW; shading indicates the $m_s = -1$ manifold.  (e-g) Fluorescence change ((signal - reference)/reference) vs RF frequency; fit shows the expected form for a RF square pulse; the resonance frequency is shown with the statistical standard deviation for the fit. 
(h-j) Fluorescence vs RF pulse duration on resonance (frequency indicated on corresponding RF spectrum) superposed on a  sinusoidal fit.
Short-time data (green) shows contrast of electron spin nutations for comparison.}
\end{figure}

Optical pumping at the excited state level-anticrossing (ESLAC) provides the polarization necessary to observe nuclear magnetic resonance (NMR). 
Working within the $m_s = -1$ manifold, we measure the nuclear spin projection as a function of both RF frequency and duration (see Fig.2), and fit our data 
to the expected form for square pulse excitation to extract the resonance frequencies $\omega_n$ and Rabi frequencies $\Omega_{n}$ for all three nuclear spin species. Observed values of $\omega_n$ are consistent with published values~\cite{Felton09}, but the Rabi frequencies $\Omega_n$ greatly exceed the naive expectation $\Omega_{n} = |\gamma_n B_{RF}| \sim$ a few kHz (we measure $B_{RF}\sim 5-10$ G). 
Indeed, the nuclear spin nutation rates are dominated by a second order process akin to a chemical shift, involving an electron spin flip 
followed by an electron-nuclear spin flip-flop, with $\Omega_{n} \approx \gamma_n B_{RF}+\frac{A_{\perp}~g \mu_B~B_{RF} }{\Delta}$~\cite{Childress06}.  
The enhancement occurs for all species of nuclear spin, enabling fast nuclear spin manipulation with low demands on RF power.


The readout mechanism for the nuclear spin nutations shown in Fig.~2 is remarkably robust, yielding an NMR contrast  
as great or greater than the ESR signal -- even for the nitrogen nuclear spins. 
Earlier nuclear spin readout schemes have relied on nuclear-spin-selective MW pulse(s)within the GS to transfer nuclear populations to measureable electron spin populations~\cite{Jelezko04b,Dutt07, Mizuochi09}.  Such readout methods do not work as well for $^{15}$N and $^{14}$N because their GS hyperfine interactions do not greatly exceed the electron spin dephasing rate.  In contrast, by working at the ESLAC we can take advantage of the larger ES hyperfine parameters (A $\sim$ 50 MHz\cite{Gali09, Fuchs08}) to directly read out the nuclear spin state:  Whereas the $m_s =0$ state with maximal nuclear spin projection fluoresces strongly, other nuclear spin projections pass through the dark singlet state during the polarization process, leading to reduced fluorescence.  The reduction in fluorescence depends on the number of passes through the dark state required to reach the polarized state, so that the average nuclear spin projection can be directly measured with a contrast as great as or even exceeding the ESR signal.  To illustrate the different fluorescence levels, Figure 3b compares an ESR spectrum with $m_I = +1$ ($^{14}$N) to one with $m_I = 0$ (created using an NMR $\pi$ pulse on the transition indicated in Fig. 3a).  The ESR spectra corroborate the nuclear spin projection and illustrate a significantly reduced fluorescence level for $m_I = 0$ as compared to $m_I = 1$.  While our measurements of this effect are aimed at direct nuclear spin readout, the same mechanism has recently been explored in great detail in~\cite{Steiner09} for applications in enhanced electron spin measurement.


\begin{figure}[htbp] 
  \centering
  \includegraphics[width=3in,height=2.5in,keepaspectratio]{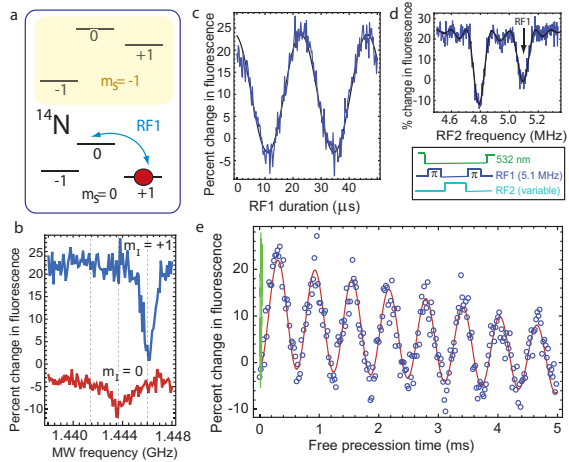}
  \label{fig:badfigure}
\caption{ (a) Energy levels for GS $m_s = 0$ $^{14}$N nuclear spin manipulation.  (b) Pulsed ESR spectra starting in $m_s =0$, $m_I = 1$ (top) or $m_s = 0$, $m_I = 0$ (bottom). Dashed lines indicate expected ESR positions for the three hyperfine levels. (c) Nuclear Rabi nutations within the $m_s = 0$ manifold (d) NMR spectrum obtained using the pulse sequence illustrated below the data. A fit to the expected square-pulse lineshape yields resonant frequencies $\omega_1 = 5.094(1)$ MHz and $\omega_2 = 4.7908(7)$ MHz.  (e) Nuclear Ramsey fringe experiment. The downward trend is likely due to electron spin relaxation, which occurs at a rate $T_1 \approx 1$ ms at room temperature~\cite{Manson06}.}
\end{figure}

The preparation, control, and detection techniques described above enable straightforward measurement of nuclear spin energy levels and coherence properties.  As an example application, we measure some of the $^{14}$N hyperfine parameters to greater precision than current values.  Fig. 3d shows direct NMR signals for $^{14}$N within the $m_s = 0$ manifold.  Combined with NMR resonant frequencies in the $m_s = 1$ manifold of $\omega = 2.9373(2)$ and $6.9580(2)$ MHz (data not shown), we find $A_{||} = -2.162(2)$ MHz and $P = -4.945(5)$ MHz, consistent with current published values~\cite{Felton09, Steiner09}.  
Coherence properties of individual nuclear spins are also readily obtained: for example, Fig. 3e shows a 2 kHz detuned Ramsey fringe experiment on the transition indicated in Fig.3a.  Although coherence properties for specific NV centers will vary with the mesoscopic distribution of $^{13}$C impurities, the long dephasing time $T_2^* \sim$ few ms observed here bodes well for $^{14}$N-based schemes. In short, these techniques open the door for complete characterization of individual nuclear spins associated with the NV center.  

Resonant RF excitation can drive transitions in arbitrary nuclear spin species in any magnetic field, but the polarization and direct readout mechanisms function only in the vicinity of the ESLAC near 510G.  This limitation is severe, especially for applications requiring optical interconnections between NV centers~\cite{Childress06PRL} or stimulated emission-depletion techniques for sub-wavelength resolution~\cite{Rittweger09}.  Near the ESLAC, any light used to excite electronic transitions can destroy information stored in nuclear spins.   
While magnetic field switching techniques developed in the context of fast field cycling NMR~\cite{Anoardo01} may allow rapid transitions between two magnetic field regimes, careful examination of the NV ES level structure indicates an experimentally simpler alternative.  

Away from the ESLAC, the electron-nuclear spin transitions that lead to optical nuclear spin polarization can be brought into resonance by microwave excitation within the ES manifold.  However, under most conditions, the dominant effect of the microwaves will simply be to flip the electron spin, circumventing the hyperfine-induced flip-flops we wish to enhance.  To induce the desired transitions, higher-order processes must prevail:  For example, the axial ($||\hat{z}$) component of the microwave magnetic field and the nonsecular hyperfine terms together induce electron-nuclear spin flip-flops through a second-order process, while transverse components ($\perp \hat{z}$) of the static and microwave fields yield similar results through a third-order process. 

\begin{figure}[htbp] 
  \centering
  \includegraphics[width=3.5in,height=4in,keepaspectratio]{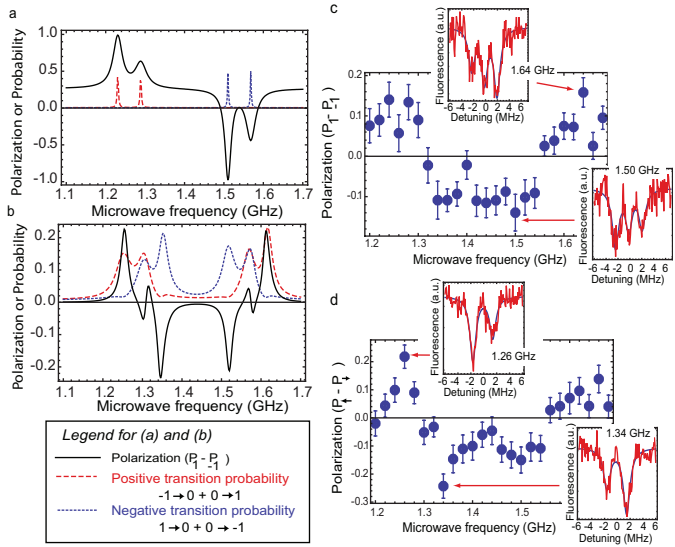}
  \label{fig:badfigure}
\caption{(a) Theoretical positive ($m_I = 0\rightarrow m_I = 1$ plus $m_I = -1 \rightarrow m_I = 0$) and negative transition probabilities and equilibrium polarization for $^{14}$N in $\mathbf{B_0} = 50\hat{z}$ G, $g \mu_B \mathbf{B_1} = 20\hat{z}$ MHz , with A = 50 MHz, P = 5 MHz, D = 1.42 GHz, $\gamma_n$ = 0.3077 kHz/G, $\gamma_e$ = 2.799 MHz/G,$\gamma = 2.5 \Gamma$, $k_{eq} = 10^{-5}\Gamma$. (b) Same as for (a), but with $g \mu_B \mathbf{B_1} = 20\hat{x}$ + 20$\hat{z}$ MHz and $\mathbf{B_0} = 40\hat{x} + 48\hat{z}$ G.  (c) Polarization of $^{14}$N in $\mathbf{B_0} = 40\hat{x} + 48\hat{z}$ G. A 2$\mu$s pulse of MW and green light is followed by a 1$\mu$s delay before probing the nuclear spin populations in the ground state with weak ESR. 
Error bars indicate one standard deviation (statistical). (d) Polarization of $^{15}$N under the same experimental conditions.}
\end{figure}

To model microwave-induced polarization, we begin with the Hamiltonian for the NV-$^{14}N$ spin system in the excited state, 
\begin{equation}
H = D S_z^2 + P I_z^2 + \left(\mathbf{B_0} + \mathbf{B_1}\cos{\omega t}\right) \cdot\left(g \mu_B \mathbf{S} - \gamma_n \mathbf{I}\right)+ A \mathbf{I}\cdot\mathbf{S},
\end{equation}
where we have allowed for static and oscillatory magnetic fields and a quadrupole splitting P, and assumed isotropic g-factors and a contact hyperfine interaction A~\cite{Gali09}.  
We then use Floquet theory~\cite{Shirley65} to calculate the probability for different nuclear-spin-changing transitions in the excited state as a function of the microwave frequency $\omega$.
In the case of $^{14}$N, these probabilities can be incorporated into a simple 3-level rate equation model because symmetry and the quadrupole moment conspire to keep the nuclear spin quantization axis $||\hat{z}$ for moderate magnetic fields.  The transition probabilities are incorporated into the model using excitation rates $\gamma$ (for transitions within the $m_s = 0$ manifold) and $\Gamma$ (for transitions between the $m_s = 0$ and $m_s = \pm 1$ manifolds).  A depolarization rate $k_{eq}$ completes the model, which we use to predict the equilibrium nuclear spin polarization $P_{+1}-P_{-1}$~\cite{Jacques09}.  
The details of this model will be published elsewhere.  
Figure 4a shows $^{14}$N nuclear spin flip probabilities and equilibrium polarization for static and oscillatory magnetic fields $\mathbf{B_0}$ and $\mathbf{B_1}$ oriented along $\hat{z}$, while Figure 4b illustrates the same quantities for off-axis magnetic fields.

The orientation of the microwave magnetic field $\mathbf{B_1}$ is constrained by the geometry of our experiments to be $\sim$45 degrees from the NV axis.  Working in a sufficiently large axial static magnetic field $B_{0,z}$ to distinguish the ES $m_s = \pm 1$ levels, with a transverse component $B_{0,x}$ large enough to enable the third-order polarization process, we apply optical excitation and microwaves before taking a GS ESR spectrum.  We fit the GS ESR signals with three (or two, for $^{15}N$) Lorentzians constrained to have the same width, and use the depth of the fits to estimate the population in each of the nuclear spin projections.  We thereby observe a small degree of polarization for both $^{14}$N and $^{15}$N nuclear spins (see Fig. 4c and d), and our results in Fig. 4c agree qualitatively with the model predictions in Fig 4b.  

Although our experiments are currently constrained by the orientation of $\mathbf{B_1}$, our model indicates that precise orientation of $B_0$ and $B_1$~\cite{Alegre07} may allow full polarization of $^{14}$N (Fig 4a). Moreover, because each polarization step involves the metastable singlet state, such a polarization technique will also provide a direct nuclear spin readout mechanism. With careful engineering, it may be possible to enjoy the robust nuclear spin polarization and readout now available at the ESLAC at arbitrary magnetic fields.   

  
In conclusion, we have demonstrated initialization, manipulation, and direct readout nuclear spins associated with the NV center in diamond, and proposed a mechanism to extend these techniques to other magnetic field regimes. Such control over nuclear spins has ramifications for a broad range of applications in quantum information science.  Moreover, these techniques enable measurements of nuclear spin parameters, interactions and coherence properties that may guide development of future quantum information processing devices.

\acknowledgements{While this manuscript was in the final stages of preparation, similar work was reported in a preprint~\cite{Steiner09}. The results reported therein corroborate many of our results, and provide a detailed characterization of the decreased fluorescence associated with ES electron-nuclear spin transitions at the ESLAC.  The authors acknowledge Yuanyuan Jiang and Gabriel Ycas for their work building and characterizing the experimental apparatus and writing software to run it.  We thank Mikhail Lukin, Jonathan Hodges, and Gurudev Dutt for valuable conversations.  This research was supported by funds from Research Corporation and from Bates College.  J.M. and B.S. acknowledge support from HHMI.}

\bibliography{paper1v2}

\end{document}